# ADVANCEMENT STUDY IN SKIN BACTERIA PROTECTION USING UV LIGHT

Ch. Jalbout†, J. Dgheim†[1]

† **Laboratory of Applied Physics, Group of Mechanical Thermal and Renewable Energies, Lebanese University**

**Abstract:** UVC radiation has been recently used to disinfect hospital, daycare and public places. It can be used to reduce bacteria by deactivating their DNA. Our work is to examine the effectiveness of UVC radiation (254 nm) in inactivating the SkinBorn bacteria. Mathematical models of these microorganisms placed in different positions on a human skin have been developed and solved numerically. Heat transfer equation is linked to Lambert Beer relationship by taking into account the conduction, the convection and the radiation phenomena. The experimental set-up is realized in order to verify our purposes. Swabs from skin and under nails of different individuals were inoculated on nutrient agar plates. Our experimental results achieved 97.43% of microorganisms killing. Both experimental and numerical results demonstrate that secured using of UVC is extremely effective in killing bacteria without damaging the human skin.

[1] Corresponding author: jdgheim@ul.edu.lb

# I. Introduction

Lately ionizing and non ionizing radiation are used to control the growth of microorganisms in clinical settings, the food industry and laboratories. Since ionizing radiation has more energy, it can penetrate cells very easily and quickly. It is used to sterilize medical supplies and some food products we eat. Only some forms of non ionizing radiation like UV light are useful for controlling microbial growth. There are three general types of ultraviolet light; UVA, UVB and UVC. Each of these types has a different wavelength. UVA and UVB have a longer wavelength than UVC; therefore, they have less energy and cannot penetrate cells as well. While they are still dangerous, they are not considered germicidal because of their small effect on microbes. The present work discusses the potential of UVC irradiation as an alternative approach to current methods used to treat bacteria, fungi and viruses without damaging the skin. A bacterium like MRSA is the most under-reported hidden danger in hospitals around the world. This bacterium has developed a resistance to many commonly used antibiotics. Although it has been known for the last 100 years that UVC radiation is highly germicidal, the use of UVC radiation for prevention and treatment of localized infections is still in the very early stages of development. Most of the studies are confined to *in vitro* levels, while *in vivo* animal studies and clinical studies are much rarer.

**Sullivan and Conner-Kerr (2000)** [1] compared the inactivation efficacies of UVC on pathogenic bacteria and fungi, in both single suspensions and mixed suspensions in vitro. The calculated irradiance at the device aperture was 15.54 $mW/cm^2$, with the distance between the UVC lamp and suspension surface is 25.4 mm. Upon exposure to UVC, a 99.9% inactivation rate was obtained at 3–5 s for the tested bacteria (P. aeruginosa and Mycobacterium abscessus). By contrast, 15–30 s of UVC treatment was required to obtain 99.9% inactivation of the tested fungi (Candida albicans, Aspergillus fumigatus).

**Thain et al. (2002)** [2] investigated the use of UVC for the treatment of cutaneous ulcer infections .Three patients with chronic ulcers infected with MRSA were treated with UVC at 254 nm. UVC irradiation was applied to each wound for 180 s, with the irradiance of 15.54 $mW/cm^2$. In all three patients, UVC treatment reduced the relative number of bacteria in wounds and facilitated wound healing. A total UVGI dose was applied for 4-8 min to the surface of the Petri dishes ranging from 12 to 96 $mJ/cm^2$. It is found that UVGI can be used to inactivate culturable fungal spores. As a result, it is found that the UVGI dose necessary to kill or inactivate 95% of the Aspergillus spp. was 35 $mJ/cm^2$for A. flavus and 98 $mJ/cm^2$ for A. fumigatus.

**Brian R.Yaun (2004)** [3] applied UVC on the surface of Red Delicious apples, leaf lettuce and tomatoes contaminated with *Salmonella* spp. or *Escherichia coli*. The samples were subjected to different doses ranging from 1.5 to 24 $mW/cm^2$ of UVC. UVC applied to apples contaminated with *E. coli* resulted in the highest bacteria reduction at 24 $mW/cm^2$. Lower reductions were seen on tomatoes inoculated with *Salmonella* spp. and green leaf lettuce contaminated with both *Salmonella* spp. and *E. coli*. No difference was seen in the ability of UVC to inactivate a higher population of either *Salmonella* spp. or *E. coli* on the surface of green leaf lettuce.

A clinical study using UVC to treat toenail onychomycosis was reported by **Boker et al. (2007)** [4]. Thirty patients with mild-to-moderate onychomycosis involving no more than 35% of the great toenail were equally randomized to receive 4 weekly UVC treatments with either a low-pressure mercury lamp delivering a total UVC dose of 22 $J/cm^2$ at the surfaces of the treated toenails or via a xenon pulsed-light device delivering a total UVC dose of 2–4 $J/cm^2$ at the surface of the treated toenails. 60% of patients treated with the xenon pulsed-light device showed an improvement of at least 1 point in their week-16. Otherwise, the patients treated with the low-pressure mercury lamp, 26% had at least a 1-point improvement in their week-16.

In vitro study was conducted by **Bhamini K. Rao (2011)** [5] to explore the effect of direct and indirect UV radiation through a transparent 0.15-mm thick transparent polythene sheet on Gram-positive cocci. Six bacterial strains were exposed to direct and filtered UVC (254 nm) for 5, 10, 15, 20, 25, and 30 seconds, then incubated for 24 hours. Direct UVC had good bactericidal effect at durations ranging from a minimum of 5 seconds to a maximum of 15 seconds. When UVC was filtered no bactericidal effect was observed. The results confirm the bactericidal effect of UVC after a short period of time.

In vitro studies realizing by **Tianhong Dai (2012)** [6] demonstrated that A. baumannii cells were inactivated at UVC exposures lower than those needed to deactivate mammalian cells. It was observed in animal studies that UVC (3.24 $J/cm^2$ for abrasions and 2.59 $J/cm^2$ for burns) significantly reduced the bacterial burdens in UVC treated wounds by approximately 10-fold compared with nontreated controls. DNA lesions were observed by immunofluorescence in mouse skin abrasions immediately after a UVC exposure of 3.24 $J/cm^2$; however, the lesions were extensively repaired within 72 hours.

In his study **David Brenner (2018**) [7], demonstrated that far UVC light can kill H1N1 viruses. H1N1 virus was released into a test chamber and exposed to very low doses of 2 $mJ/cm^2$ of 222 nm far UVC light. The far UVC light efficiently inactivated >95% of aerosolized H1N1 influenza virus, with about the same efficiency as conventional germicidal UV light.

This research consists on improving a prototype and the numerical model of heat and radiation equations of the Bacteria treatment on the skin. The equation will be solved by using numerical software. Initial and Boundary conditions will be proposed in order to determine the temperature evolution on the bacterium and the skin. On the other hand the experimental set-up is improved in order to verify our purposes. Swabs from skin and under nails of different individuals were inoculated on nutrient agar plates and exposed to UVC radiation.

In part *II*, the mathematical models describing different types of bacteria placed in different positions on the human skin are studied and solved numerically. In part *III*, the numerical results are presented under the form of temperature profiles. In part *VI*, the experimental set-up is performed to test the efficiency of UVC on SkinBorn bacteria.

# II Mathematical modeling

## *II.1 Physical Model*

The physical model is formed from a UVC lamp that emits UVC radiation on the stratum cornium. The physical phenomenon is described by the bioheat equation linked to Lambert-beer's law. In the aim to sterilize the skin tissue from the bacteria with a maximum short time in few seconds, a determined UVC fluence is sent to the bacteria that will be killed at an average temperature of $48^0$C (321K) without damaging the skin tissue (figures 1 & 2). On the skin surface, several bacteria are placed and exposed to a UVC lamp of wavelength $\lambda$=254 nm, a length $L$=0.5 m, a diameter $D$=2.5 $10^{-2}$ m, and an electrical power $P$=6 W. The distance lamp-skin is $d_1$=7 $10^{-2}$ m and the centered distance between the lamp and the hand is $R=d_1+(D/2)$. That is equal to 8.25 $10^{-2}$ m. The bacteria are mostly composed from water of imaginary index of refraction $n_2$=1.25 $10^{-7}$.

Table 1 lists the physical properties of the different mediums:

| Property | Bacteria (Spherical, Rod-shaped) | Skin tissue |
|---|---|---|
| Density, $\rho$ (kg/m$^3$) | 980 | *1100* |
| Thermal conductivity, *K* (W/(m.K)) | *0.7* | *0.23* |
| Specific heat, *C* (J/kg.K) | *4100* | *3300* |

***Table 1:*** *The physical properties of the different mediums*

Figure 1 show five Rod-shaped bacteria placed above each other. The stratum cornium where these bacteria are placed has a dimension of 10x10 µm$^2$. These bacteria are exposed to the same UVC fluence.

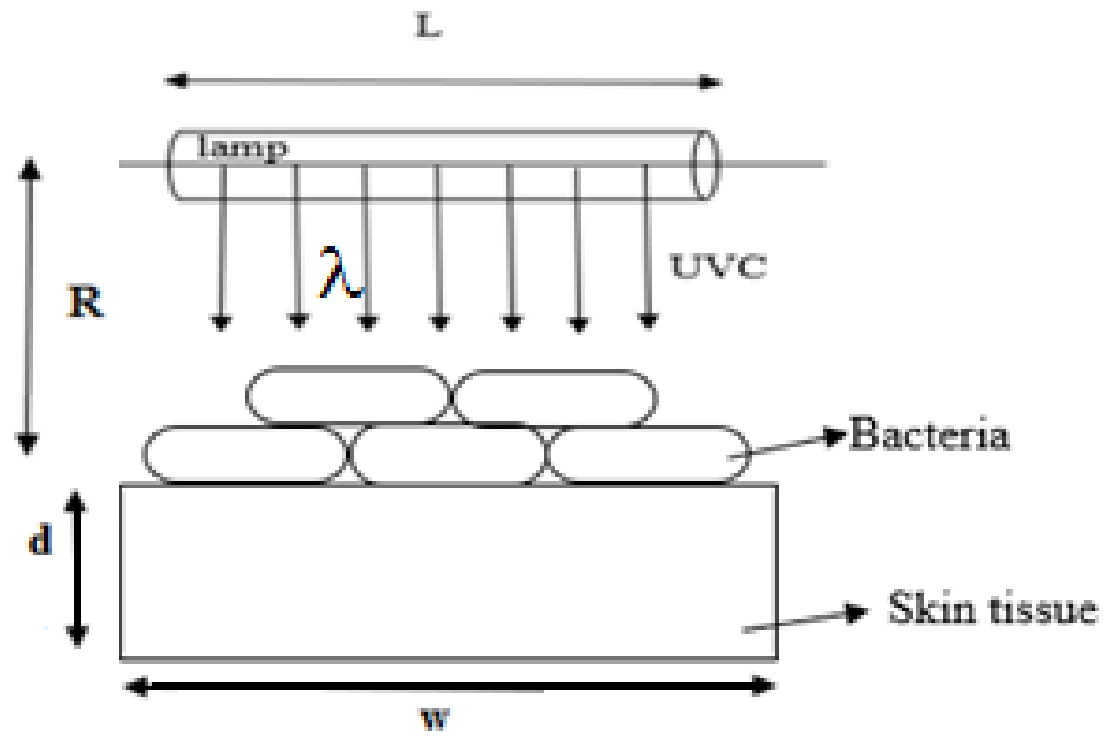

***Figure 1:*** *Two rows of Rod-shaped bacteria exposed to UVC radiation*

Figure 2 shows six spherical bacteria placed above each other. Each one has a diameter of 0.1 µm. The stratum cornium where these bacteria are placed has a dimension of 0.6x0.6 µm$^2$. The temperature field is computed for the sample, for the border of heated and unheated region.

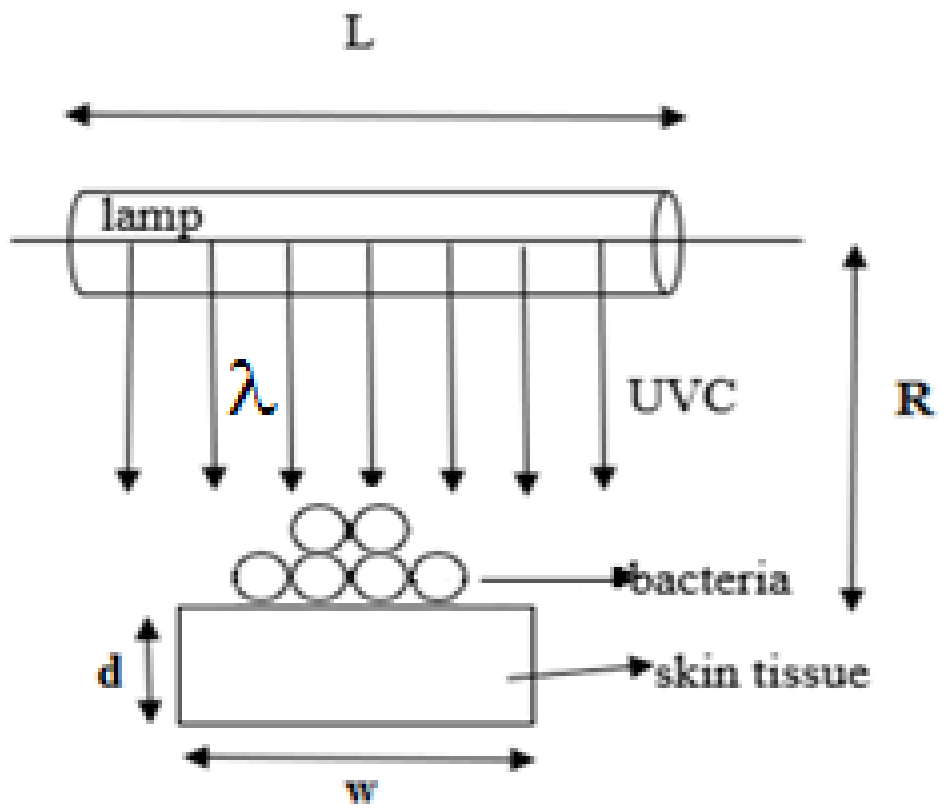

***Figure 2:*** *Two rows of spherical bacteria exposed to UVC radiation*

## *II.2 Mathematical equations*

The mathematical model is built in Cartesian coordinate where (*O*) is the origin of the system and (*y*) counted positively toward the upper side of the system. Both physical models have the same mathematical equations.

The bioheat equation is coupled to the radiation equation by the continuity of the conduction and the convection at the interface. In two dimensions and in the (*Oxy*) coordinates, the heat equation is written as the following:

$$\rho C \frac{\partial T}{\partial t} = \frac{\partial (K\frac{\partial T}{\partial x})}{\partial x} + \frac{\partial (K\frac{\partial T}{\partial y})}{\partial y} + Q_{ext} \qquad (1)$$

Where $T$ the temperature, t is the time and $Q_{ext}$ is the UVC volumetric power.

Since the UVC radiation is directed toward the bacteria without damaging the skin tissue, the blood perfusion and the metabolism heat source are considered negligible. The UVC heat equation that depends on Lambert-beer law is written as the following:

$$Q_{ext} = q_0 a e^{ay} \qquad (2)$$

Where $q_0$ is the UVC surface power and *a* is the absorptivity.

*The UVC* surface power *is given by:*

$$q_0 = I_0/\tau \qquad (3)$$

Where $I_0$ is the UVC fluence and $\tau$ is the UVC exposure time.

The absorptivity formula is given by:

$$a = \frac{4\pi n_2}{\lambda} \qquad (4)$$

### *II.5 Initial and boundary conditions*

To complete the mathematical models, the initial and boundary conditions are given.

#### *II.5.1 Two rows of Rod-shaped bacteria*

For the two rows of rod-shaped bacteria (figure 3):

- The Initial condition for $t<t_0$ : $T=T_0$.
- The Boundary conditions for $t>t_0$:

At the green boundaries, $T=T_0$.

At the boundaries that are exposed directly to UVC (red boundaries), the continuity of the conduction, convection and radiation is taken in consideration:

$$\vec{n}(K\nabla T) = q_0 + h(T_{inf} - T) \qquad (5)$$

Where $\vec{n}$ is the normal vector, $h$ is the convection coefficient of the ambient medium (the air), and $T_{\text{inf}}$ is the ambient medium temperature.

At the black boundaries, the continuity of the conduction is applied:

$$\vec{n}.\,(k_u \nabla T_u - k_d \nabla T_d) = 0 \qquad (6)$$

Where, $k_u$ and $T_u$ are the thermal conductivity and the temperature of the skin respectively. $k_d$ and $T_d$ are the thermal conductivity and the temperature of the bacteria respectively. For the conduction continuity between the bacteria each other, one can take $k_u = k_d$.

At the turquoise boundaries, the heat flux is given by the continuity equation of the conduction and the convection:

$$\vec{n}.(K\nabla T) = h(T_{\text{inf}} - T) \qquad (7)$$

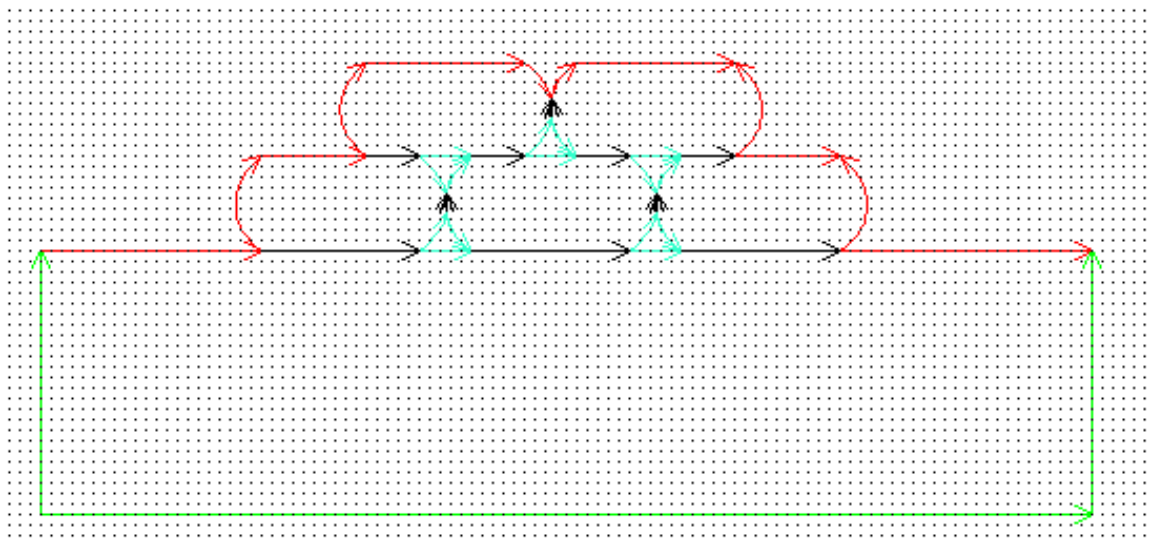

***Figure 3:*** *Sketch of the rod-shaped model showing the boundary conditions*

#### *II.5.2 Two rows of spherical bacteria*

For the two rows of spherical bacteria (figure 4), the same conditions are taken as above.

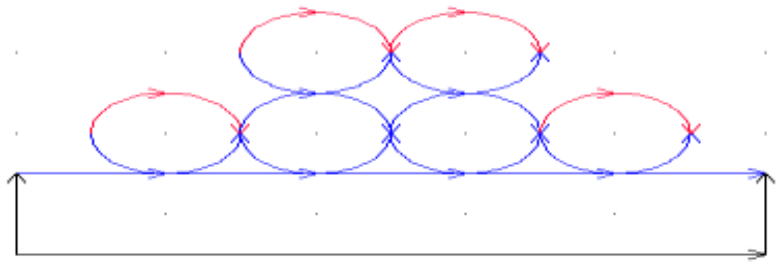

***Figure 4:*** *Sketch of the spherical model showing the boundary conditions*

## III. Numerical results and discussion

The mathematical model is solved numerically according to Dgheim et *al* 2015 [8].

The UVC in hospitals is used to provide rapid and effective disinfection of patient-related equipment using a fluence of 90 J/m$^2$ to achieve high level bacterial and viral disinfection. In our case, the human hand is exposed to a UVC irradiance of 3.18 J/m$^2$. In the case of a higher radiation dosage, UVC radiation causes red skin (erythema) or ulceration and severe acute damage to the eye of humans (photokeratitis, conjunctivitis). This is why the threshold value of 30 J/m² radiation dosage for 8 hours daily must not be exceeded as reported by the EU directive 2006.

From thermal point of view, our models assume that heat occurs only in the bacteria without damaging the skin tissue. The results of our numerical models are presented under surface and linear temperatures evolution.

### *III.2 Two rows of Rod-shaped bacteria*

The thermal contact between the first row of bacteria and the skin is realized by the conduction continuity raising the temperature of the skin surface. The conduction continuity takes place also between the two rows and increases the temperature of the bacteria of the first row allowing their kill after a little time. The convection phenomenon occurring in the air reduces the heat transfer and decreases the temperature in the skin and bacteria surrounding by the air *(figures 5)*.

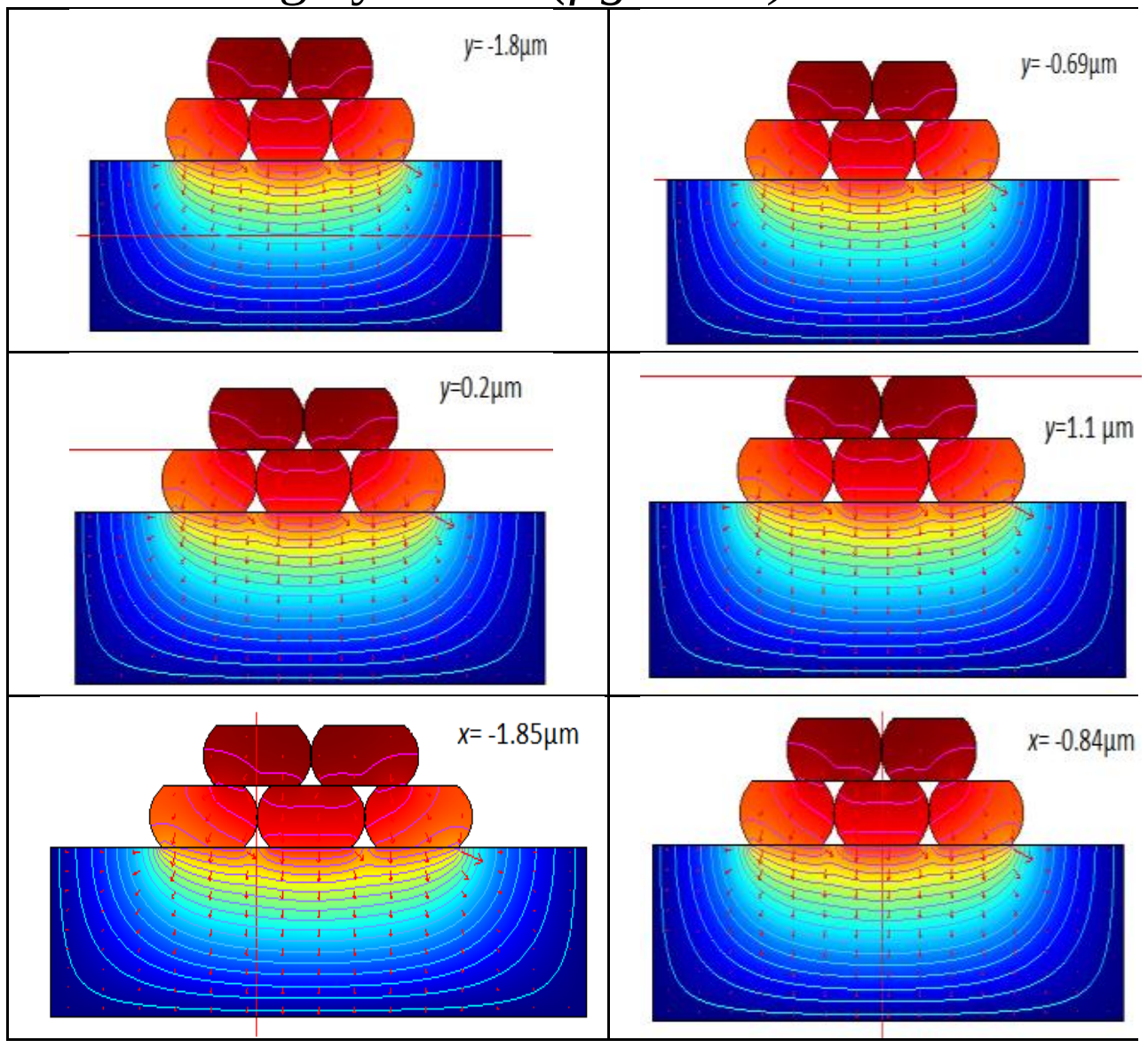

***Figure 5:*** *The heat transfer model of Rod-shaped bacteria placed in two rows*

Figure 6 shows the horizontal evolution of the temperature at the middle of the skin tissue ($y$= -1.8µm) for different times. The maximum value reached by the temperature is about 315 K under the middle bacterium.

Figure 7 shows the horizontal evolution at the skin surface ($y$=-0.69µm) for different times. The thermal contact between the first row of bacteria and the skin is realized by the conduction continuity. That's why the temperature of the skin surface increases to 320 K under the middle bacterium. Also, the reduction of temperature at $x$= -2 $10^{-6}$m and $x$= 0 m is due to the air gaps between the bacteria.

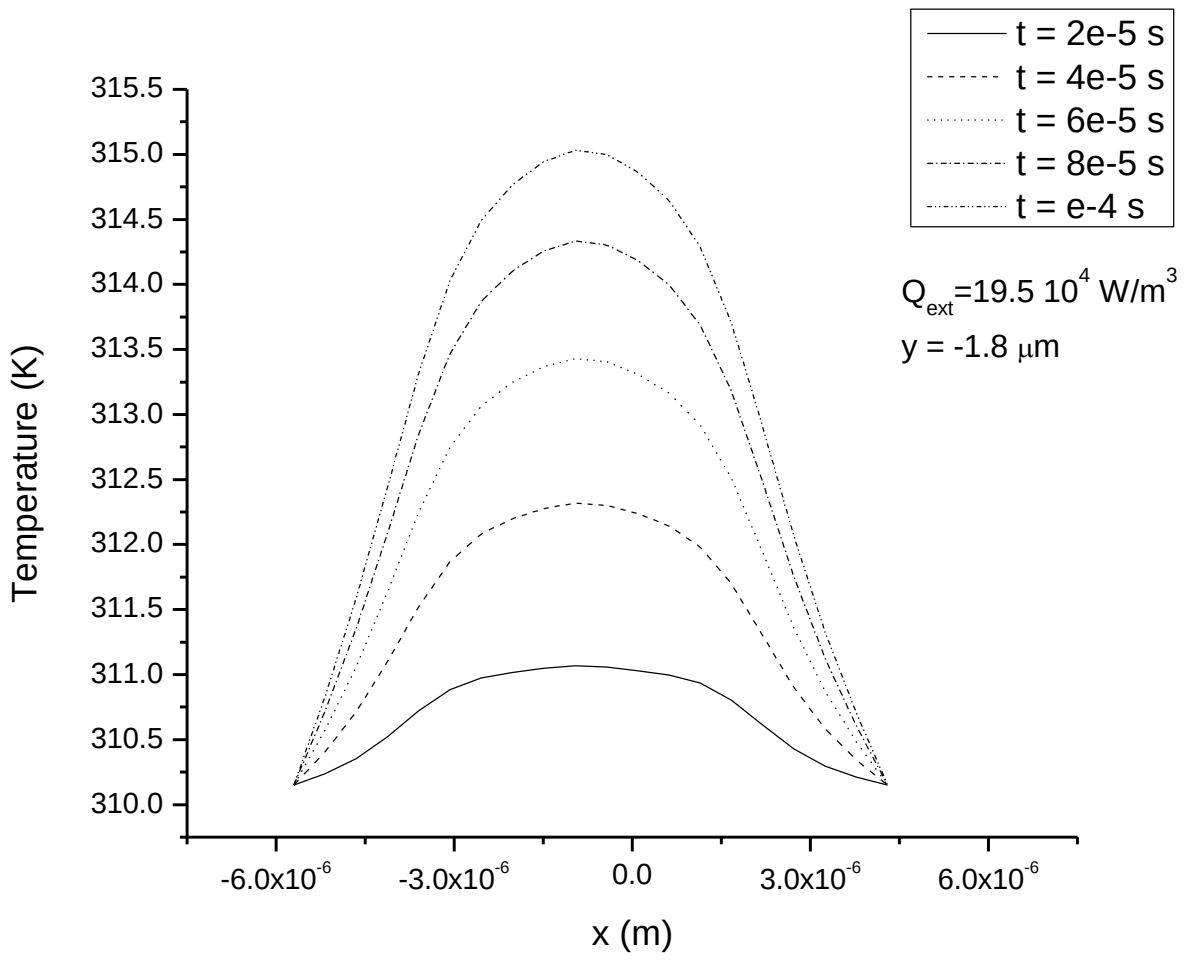

***Figure 6:*** *Temperature evolution versus x axis for different times in the middle of the skin*

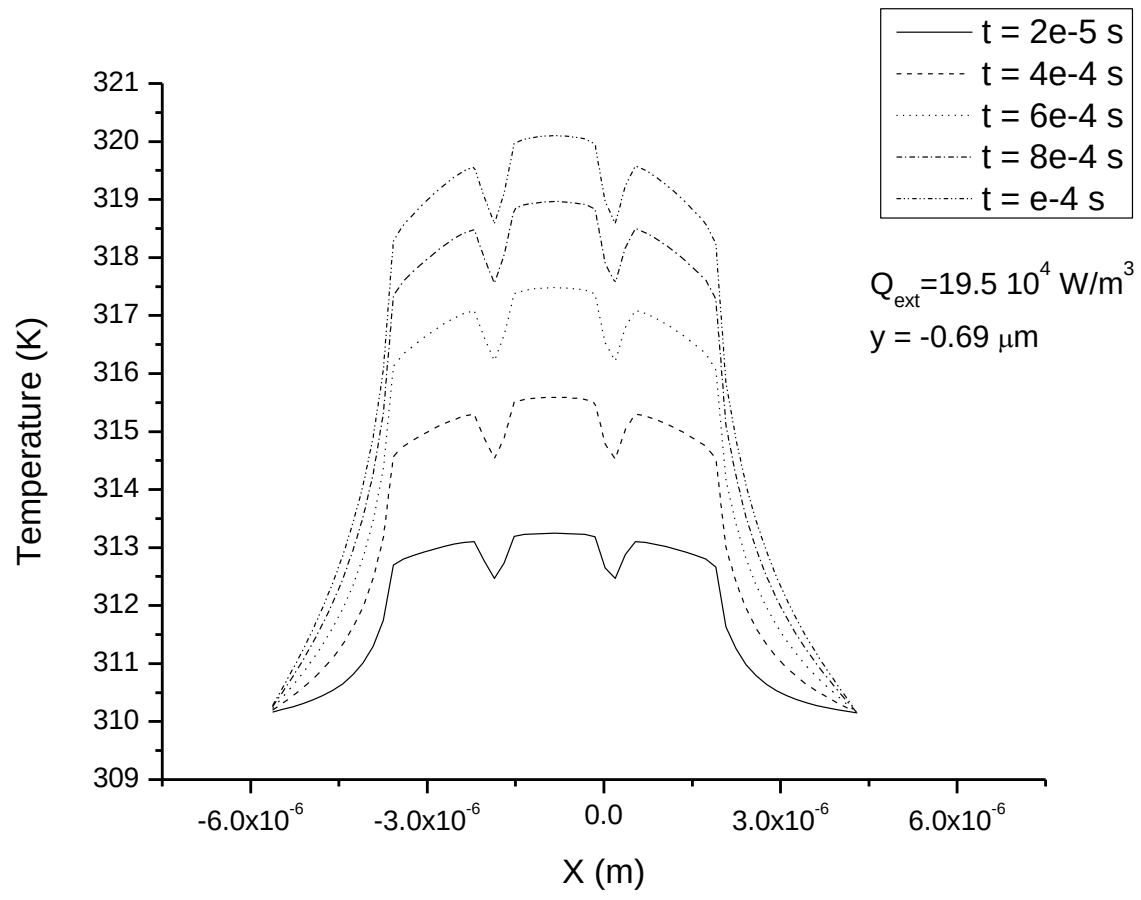

***Figure 7:*** *Temperature evolution versus x axis for different times at the skin surface*

Figure 8 shows the horizontal evolution at the first row of bacteria ($y$=0.2µm) for different times. High value of temperature is reached in the middle bacterium comparing to the border ones because of the convection in the ambient medium that reduces the temperature gradient in the left and right sides of the bacteria. The conduction continuity between the second and first row of bacteria leads to the augmentation of the temperature in the middle bacterium.

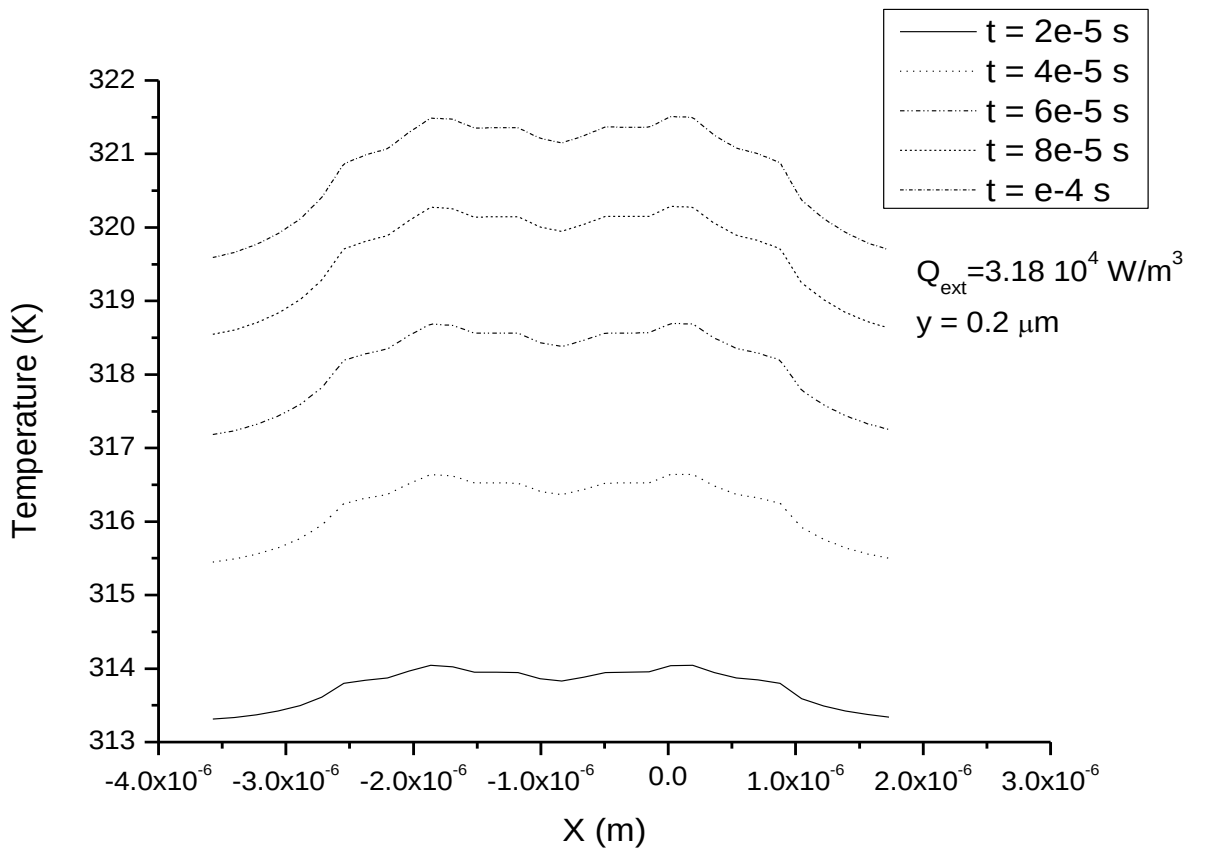

***Figure 8:*** *Temperature evolution versus x axis for different times at the first row of bacteria*

Figure 9 shows the horizontal evolution at the second row of bacteria (*y*=1.1 µm) for different times. The temperature is approximately constant (321.5 K) which means that the second row is totally exposed to UVC and therefore the bacteria reached the killing temperature (321.5 K).

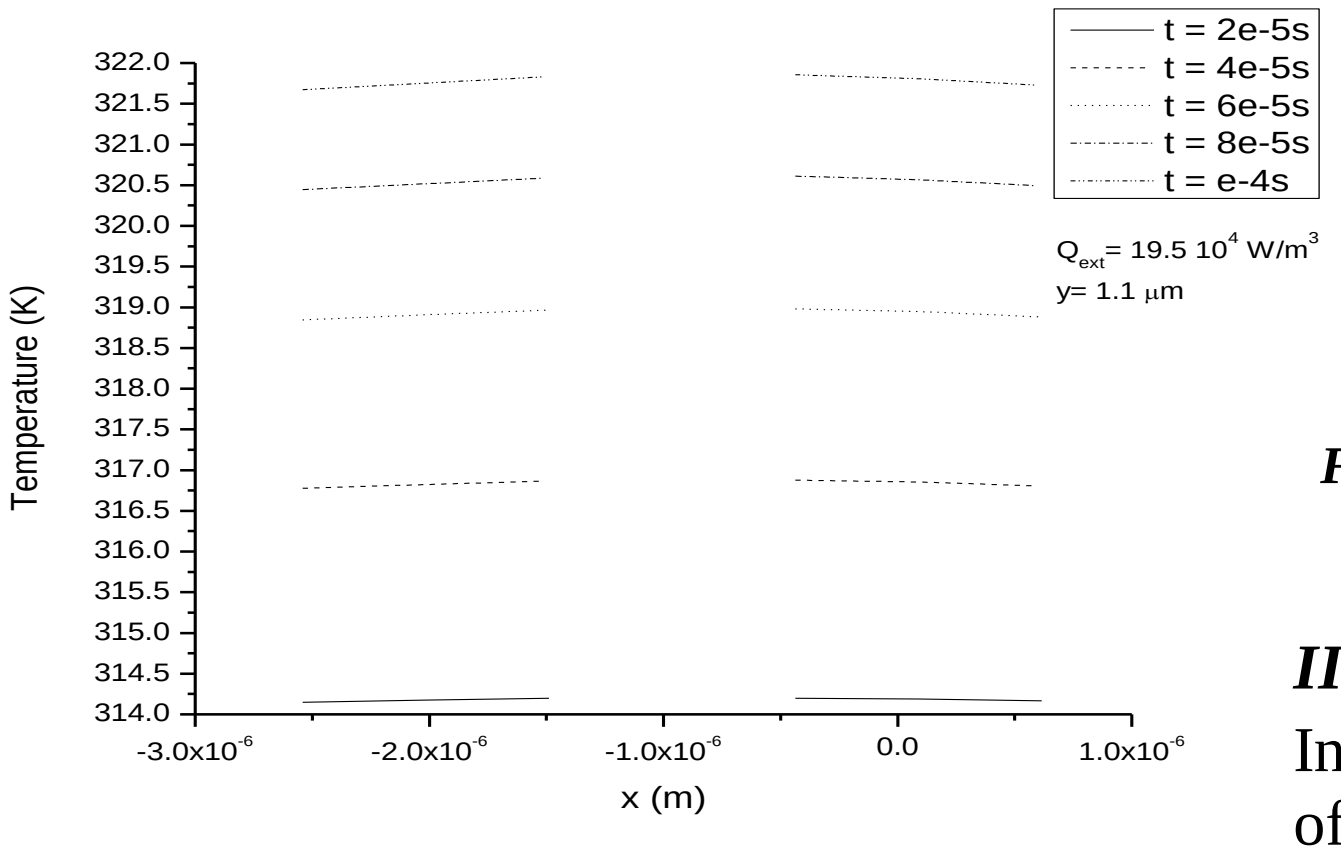

***Figure 9:*** *Temperature evolution versus x axis for different times at the second row of bacteria*

Figures 10 and 11 show the vertical evolution of the temperature at *x*= -1.85 µm and *x*= -0.84 µm respectively for different times. The temperature reaches 321.5 K at the second row surface for the two positions of *x* axis. The break line is due to the air gaps presented between the bacteria.

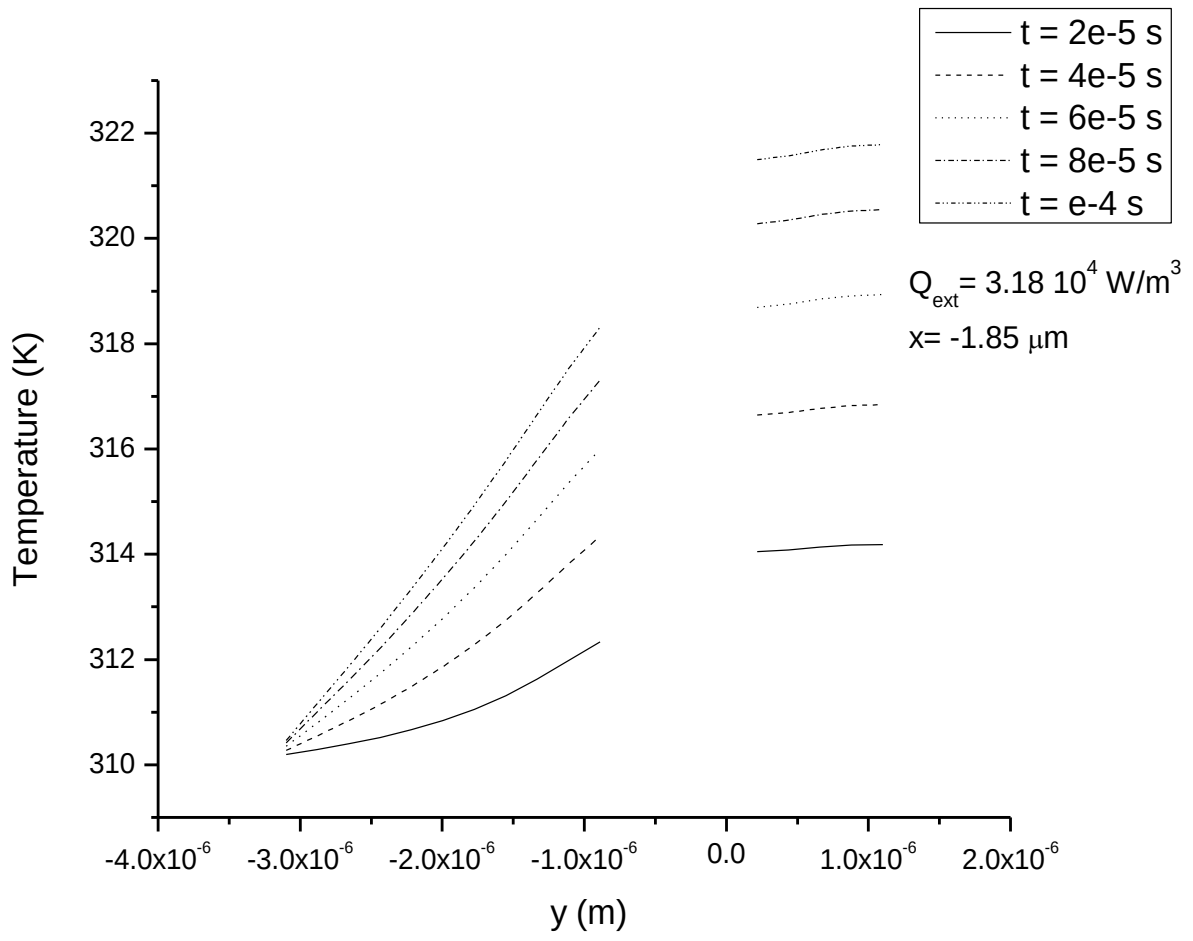

***Figure 10:*** *Temperature evolution versus y axis for different times at x=-1.85µm*

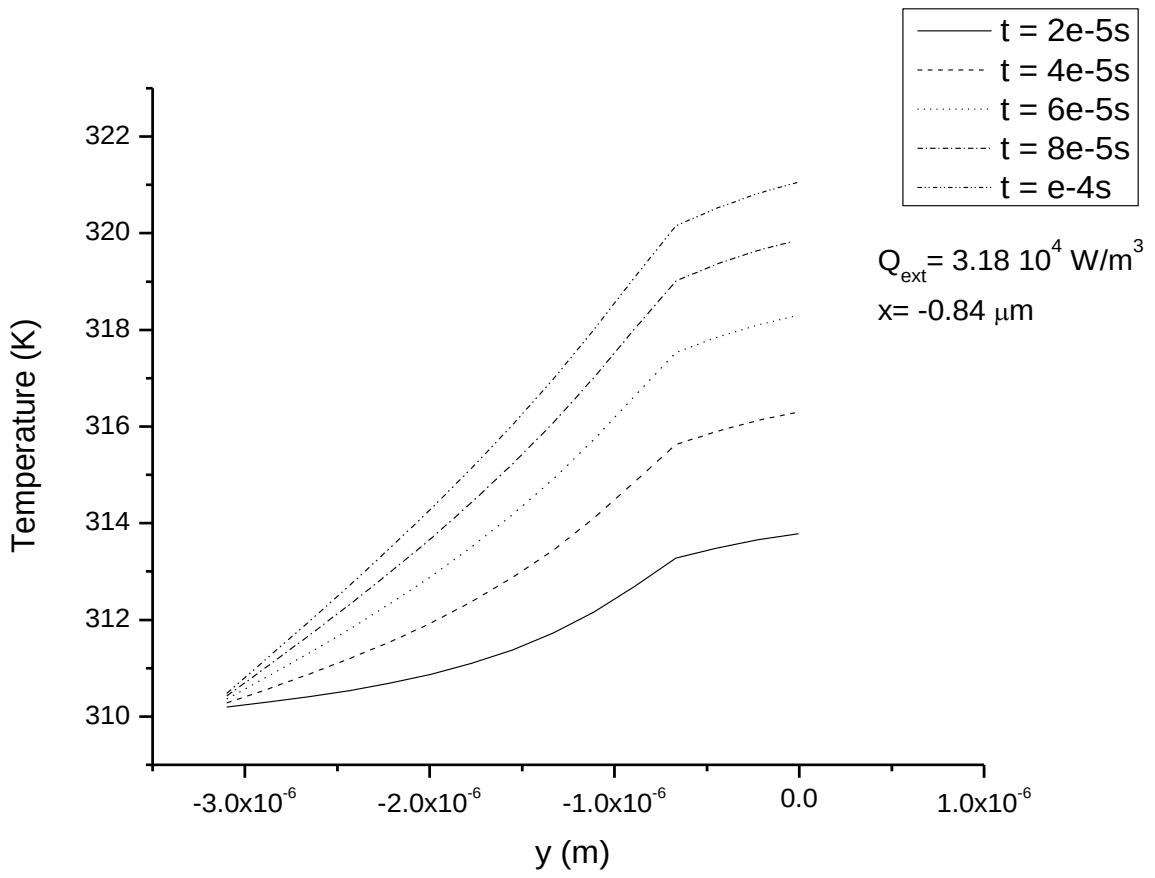

***Figure 11:*** *Temperature evolution versus y axis for different times at x= -0.84µm*

### *III.3 Two rows of spherical Bacteria*

In spherical bacteria and from thermal point of view, the thermal conductivity improves the evolution of the temperature that increases to reach a maximum value in order to kill the bacterium. At the bacteria-skin contact, the conduction continuity leads to the transfer of heat from the bacteria to the skin. The augmentation of temperature at the skin surface is shown by the isotherms that form circular cells. Moreover, the convection phenomenon due to the presence of air between bacteria and skin slows down the heat transfer and decreases the temperature in the skin.

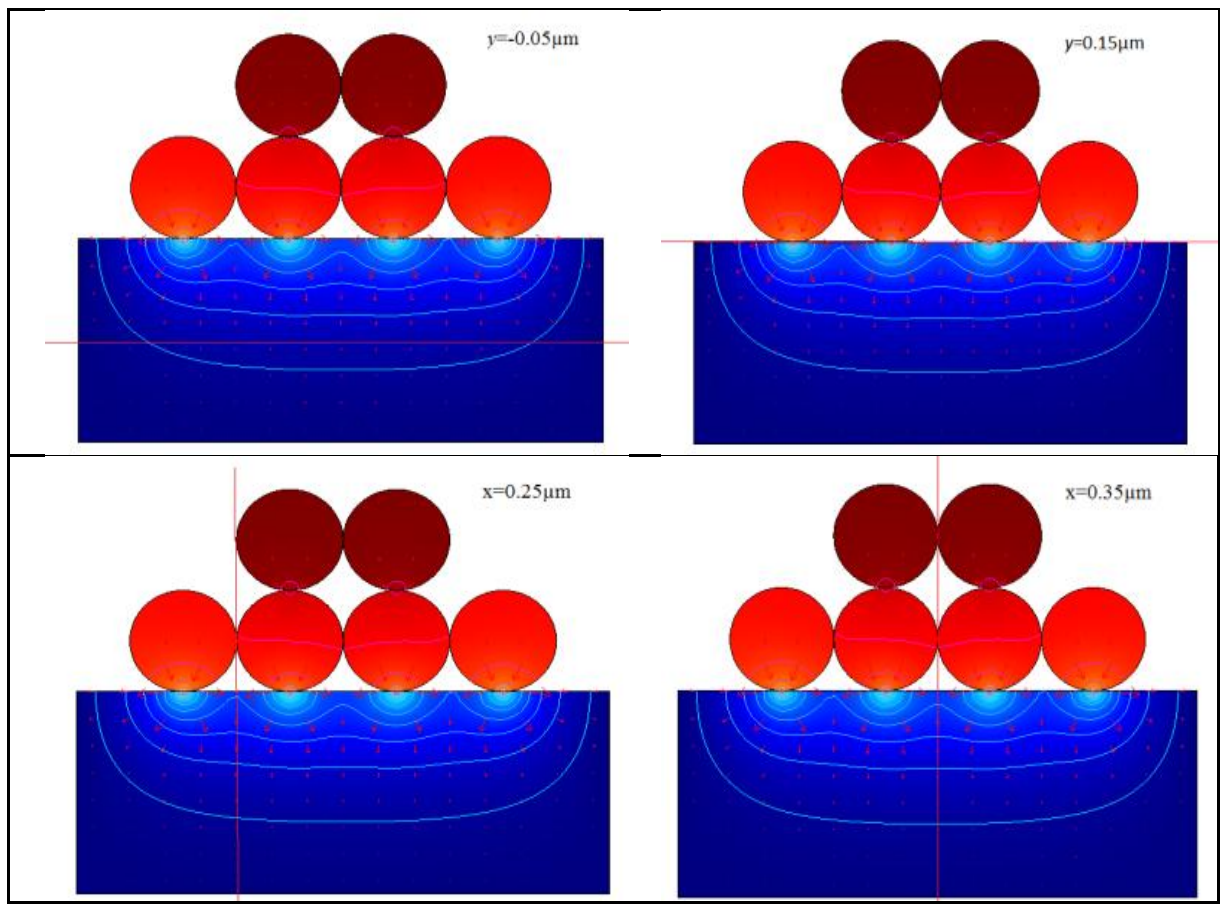

***Figure 12:*** *The heat transfer model of two rows spherical bacteria*

Figure 13 shows the temperature evolution at the skin surface ($y$=0.15µm). The temperature reaches 314.9 K at $x$=4.9 $10^{-7}$ m and then returns to 310.15 K at the edge of the surface. The augmentation of temperature by approaching the skin surface is due to the conduction continuity between the bacteria and the skin. The small reduction of the temperature between the above increased values signifies the presence of convection phenomenon in the air gaps existing between the bacteria and the skin.

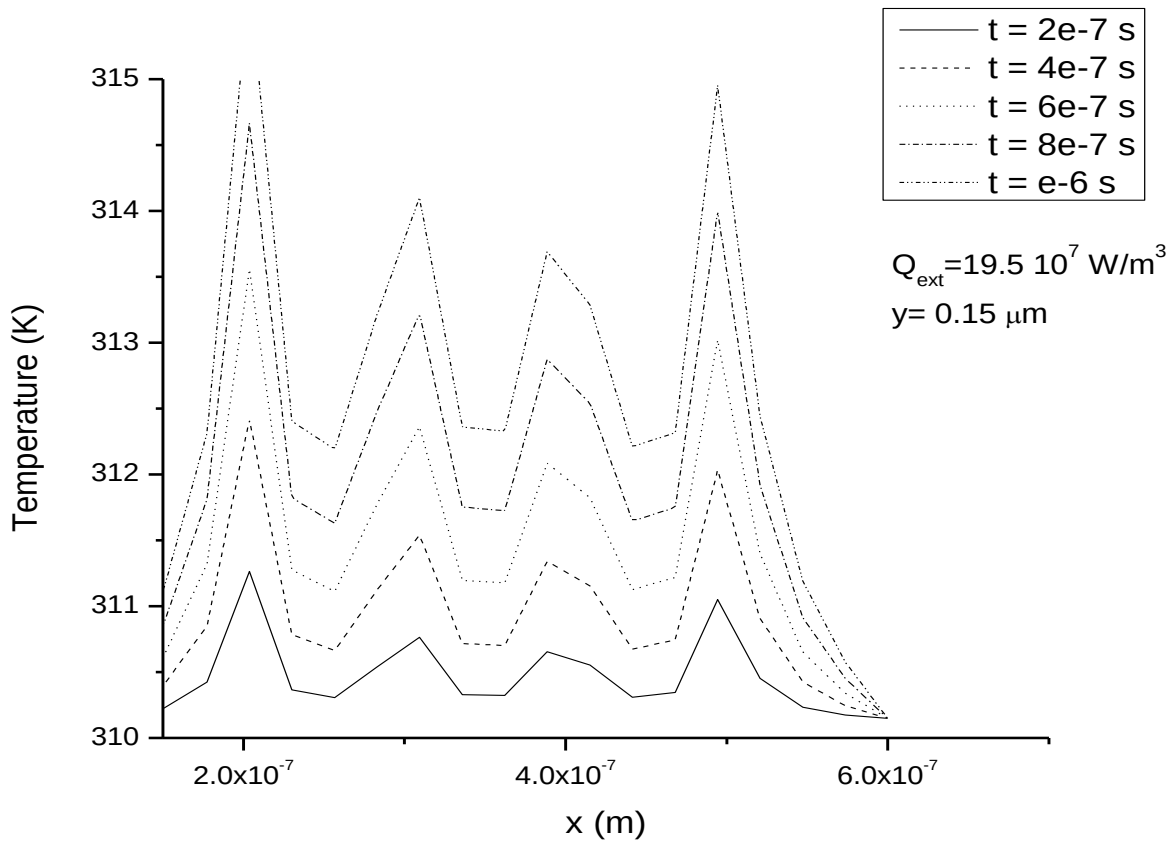

***Figure 13:*** *Temperature evolution versus x axis for different times at the skin surface*

The vertical evolution of the temperature profile is represented in figures 14 and 15 for different times. The same phenomenon is observed as the above explanation.

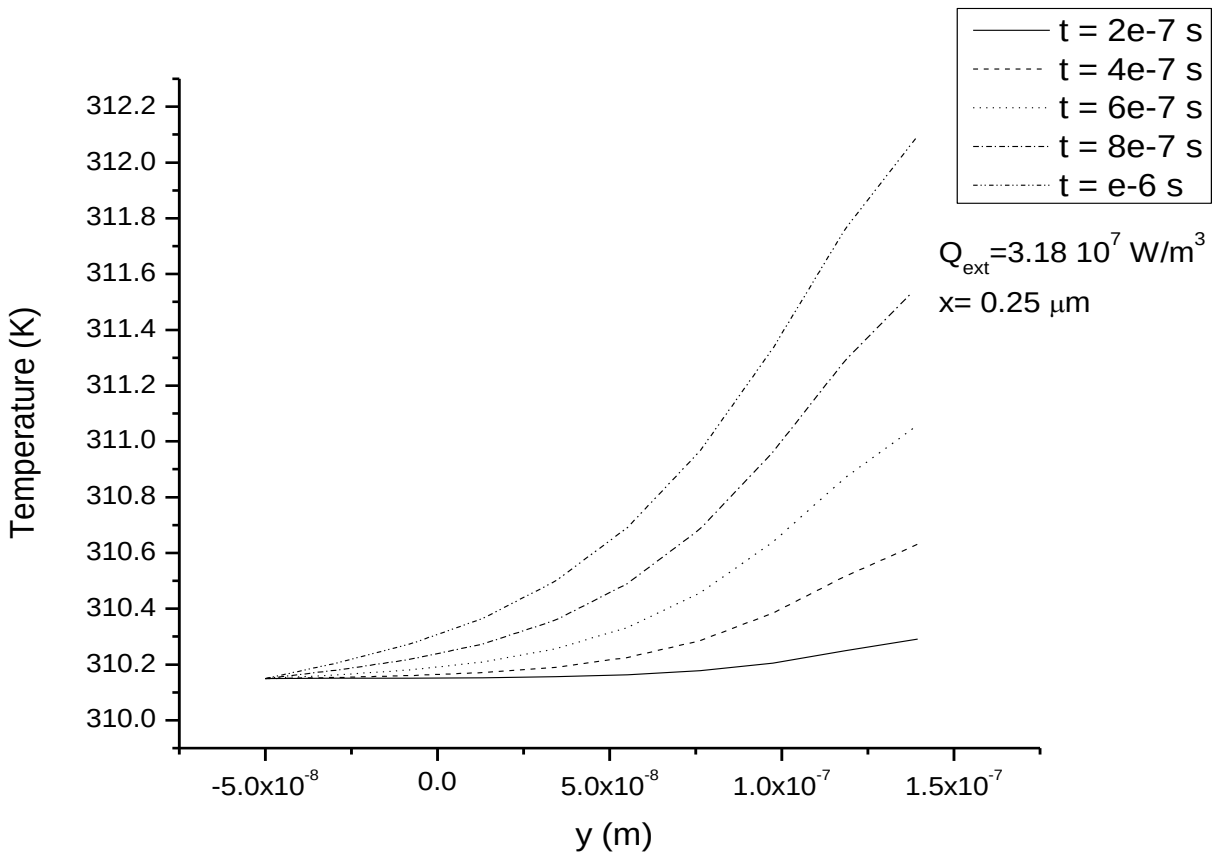

***Figure 14:*** *Temperature evolution versus y axis for different times at x=0.25µm*

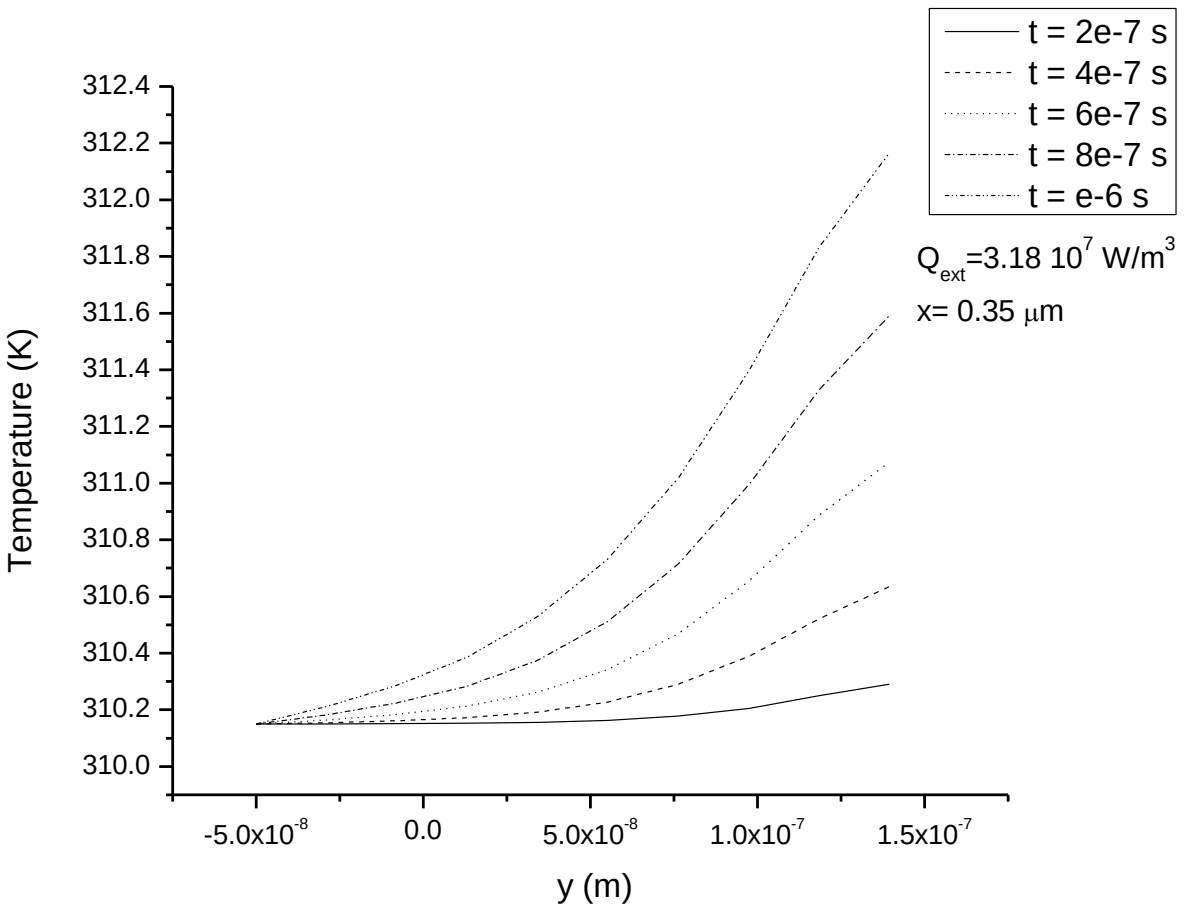

***Figure 15:*** *Temperature evolution versus y axis for different times at x=0.35µm*

# IV. Experimental results and discussion

Our device uses a UVC lamp with an average hand exposure of 10 seconds. ISO 15858:2016 and European Directive 2006 EU are taken into consideration.

## *IV.1 Preparation of the nutrient agar medium*

The preparation of the nutrient agar medium follows the following steps:

- Suspend 20 g of nutrient agar 2% (BioKar) powder in 1 liter of distilled water.
- Heat this mixture while stirring to fully dissolve all components.
- Allow the mixture to cool but not solidify.
- Pour nutrient agar into each plate and leave plates on the sterile surface until the agar has solidified.

### *IV.2 Preparation of the samples*

In order to prepare the samples one can follow these steps:

- Using a sterile swab rub on: hand, under the nails.
- Inoculate the sample on the nutrient agar prepared previously.
- Cover half of the Petri dish with a sheet of paper.
- Expose the different samples taken for 10 s to the UVC radiation.
- Incubate Petri dishes for 48 hours at 37 °C.
- Observe the radiated part and compare it to the other.

### *IV.3 Our experimental results*

After several trials a number of results are obtained. Some of the cultures were classified as uncountable *(figure 16)* because of the proliferation of the bacteria present in the medium. But one can note that these results are considered positive because the exposed area was 100% clean. On the other hand, other cultures presented countable colonies *(figure 17)* and were used to determine the survival ratio after the exposure to UVC. The results are presented in figures 16-18.

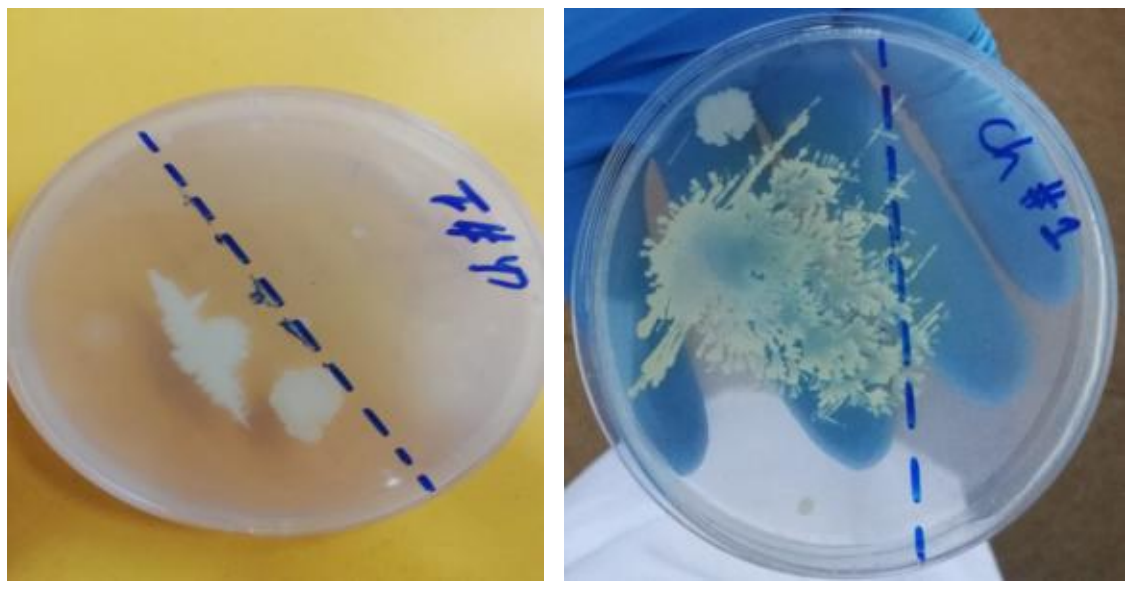

**Figure 16:** *Uncountable results*

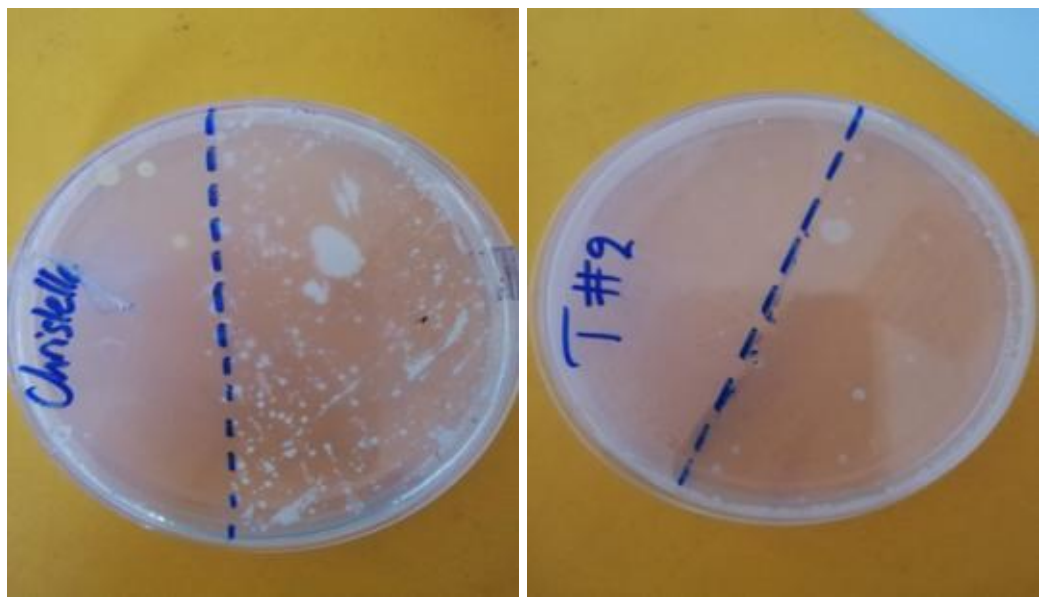

**Figure 17:** *Countable results*

One can note that after warming up the lamp the survival rate decreased in a remarkable way to reach values between 5% and 0.63% in the other samples.

In addition to that, the survival ratio of all samples including the first and second try before warming up of the lamp has a value of 2.57%. This means that the killing ratio is equal to 97.43% which is considered to be high.

The number of survived colonies is given by the following histogram:

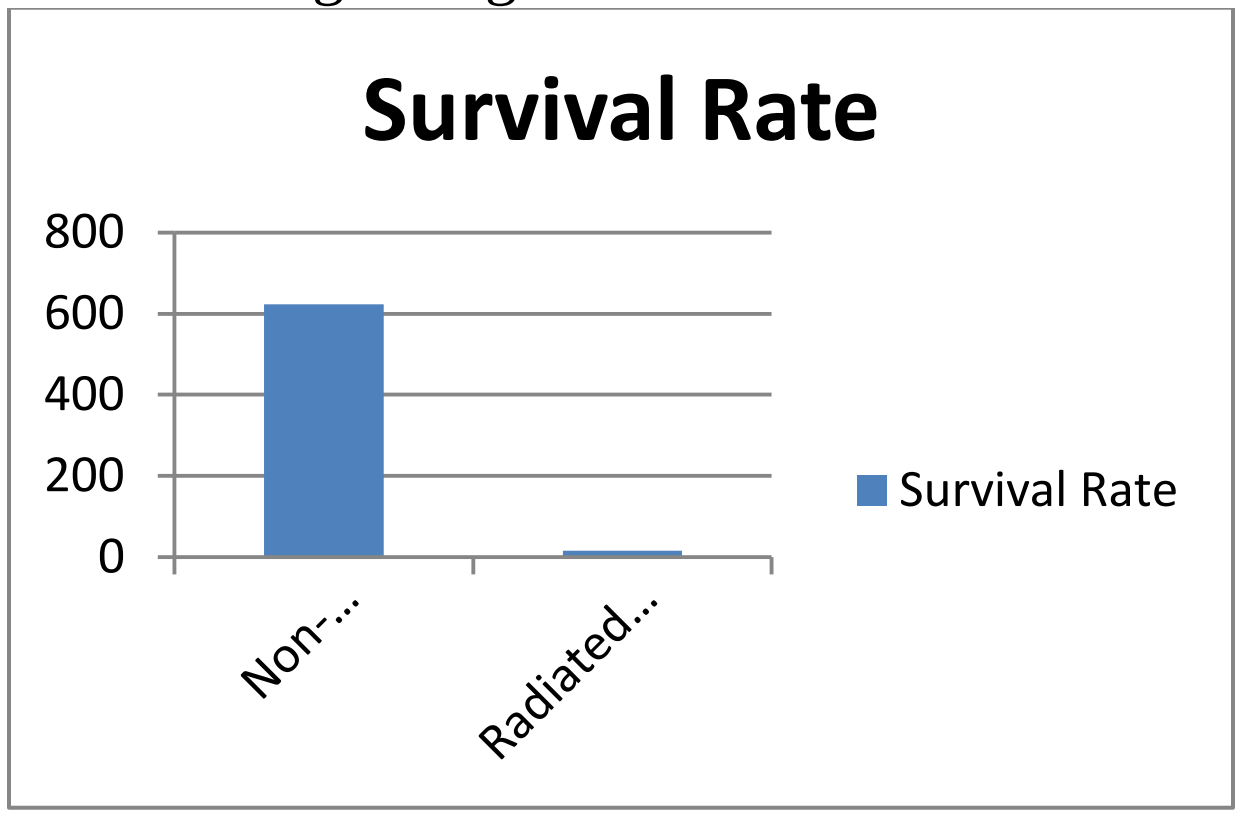

**Figure 18:** *The number of survived colonies*

# V. General conclusion

Our work aims to disinfect SkinBorn bacteria. It was divided into two parts: a numerical part and an experimental one. Two physical models concerning rod-shaped and spherical bacteria are studied. Then, the mathematical model that links the heat transfer equation to Lambert-Beer law is developed for each case. Each model is completed with the simplified conditions and the initial and boundary conditions. From thermal point of view, the numerical results are presented under the form of temperature profiles. The models consider a UVC irradiance of 3.18 J/m$^2$ for different times. As the time increases, the temperature profile also increases. Thus, our fluence of UVC radiation is capable of killing the SkinBorn bacteria, by increasing their temperature to 321 K without damaging the human skin. The UVC radiation heats up the two rows of spherical bacteria easily and in a short time compared to the two rows of rod-shaped bacteria.

From the experimental point of view, samples of the bacteria found on hands of different individuals were inoculated in a special media, irradiated then incubated for 48h. The results showed that the UVC lamp and the time of exposure were efficient in killing 97.43% of SkinBorn bacteria. Keeping in mind that warming up the lamp is an important key factor to achieve high killing percentage. From a microbiological perspective, the UVC light has proved this germicidal effect on the bacteria of the human skin.

One can prove that the dose of UVC supplied by the lamp is under the European norms (30 J/m$^2$), as well as the American ones (60 J/m$^2$), and respecting the safety requirements of the ISO 15858:2016.